\documentclass[runningheads]{llncs}
\usepackage[T1]{fontenc}
\usepackage{graphicx}
\usepackage{cite}

\begin{document}
\title{Neuro-Symbolic Artificial Intelligence for Patient Monitoring}
%
%
\author{Ole Fenske\orcidID{0009-0001-3055-478X} \and
Sebastian Bader\orcidID{0000-0001-8786-6242} \and
Thomas Kirste\orcidID{0000-0002-8956-9820}}
\authorrunning{O. Fenske et al.}
%
\institute{Rostock University, Albert-Einstein-Str. 22, 18059 Rostock, Germany
\email{\{ole.fenske,sebastian.bader,thomas.kirste\}@uni-rostock.de}\\
\url{https://www.mmis.informatik.uni-rostock.de/en/}}
\maketitle              
\begin{abstract}
  In this paper we argue that Neuro-Symbolic AI (NeSy-AI) should be applied for patient monitoring. In this context, we introduce patient monitoring as a special case of Human Activity Recognition and derive concrete requirements for this application area. We then present a process architecture and discuss why NeSy-AI should be applied for patient monitoring. To further support our argumentation, we show how NeSy-AI can help to overcome certain technical challenges that arise from this application area.    

\end{abstract}
%
%
%
\section{Introduction}
\label{sec:NeSy}

The intuition behind Neuro-Symbolic AI (NeSy-AI) is to integrate low-level perception with high-level reasoning by utilising (deep) neural networks (NNs) and symbolic systems respectively \cite{hitzler_neuro-symbolic_2022, sarker_neuro-symbolic_2021}. Here, the symbolic systems refer to explicit knowledge representation (e.g., graphs, ontologies, formal logic) and symbol manipulation algorithms. 
There are three existing approaches to realize NeSy-AI:
\begin{itemize}
    \item \textbf{Extraction}: Learn from raw data with a neural network and extract the learned information to model it as symbolic knowledge.
    \item \textbf{Embedding}: Embed symbolic knowledge right into the neural network.
    \item \textbf{Hybrid}: Combine both methods (neural and symbolic) such that they exist as co-routines next to each other.
\end{itemize}
The intuition behind such an integration is to complement both methods in their advantages and shortcomings. For instance, neural networks, on one hand, are proven to learn from raw data and are able to tackle noise and inconsistencies. Thus, applications requiring raw data analysis and information extraction can benefit from neural networks. However, the blackbox nature, i.e., inability to interpret the decision making process and learned knowledge, of such networks can be problematic in many applications such as healthcare. 

On the other hand, symbolic AI is interpretable by design where the decision process is transparent and accessible for humans. The transparent nature of symbolic AI can therefore be useful in complementing the mentioned blackbox nature of neural networks. Nevertheless, symbolic AI also suffers from certain drawbacks. For instance, processing raw data is problematic as such methods are prone to observation errors, inconsistencies and data outliers. Moreover, training symbolic AI systems, as we do with neural networks, is much harder.

The contributions of this paper are: In Section \ref{sec:patient_monitoring} we introduce patient monitoring as an important use case within Human Activity Recognition and several requirements are provided.
In section 3 we derive a process architecture for patient monitoring and justify our intuition behind applying Nesy-AI. Furthermore, we outline the concrete challenges to be expected and strategies to tackle these challenges through Nesy-AI on an abstract and theoretical level. Finally, in the last section, we conclude our work by summarizing the core insights of this paper and giving a short outlook. 

\section{Patient monitoring as a special case of Human Activity Recognition}
\label{sec:patient_monitoring}
Human Activity Recognition (HAR) aims at recognizing activities of interest based on sensor information \cite{jaeyoung_yang_activity_2011}. 
Depending upon the type of sensor used, HAR can be classified into vision and sensor based \cite{hussain_different_2020}. Vision based approaches rely on cameras whereas sensor based methods can be further divided into wearable sensors and environmental sensors. Wearable sensors are often attached to humans or objects that are part of the activity i.e. cooking utensils, whereas environmental sensor are placed in the surroundings.

In this context, patient monitoring can be seen as a special case of HAR, because we want to monitor the current state at a patients bedside (e. g. nurse and patient next to bed) as well as specific high-level activities (e.g. changing bedsheets) in hospitals. Therefore, we deploy non-invasive sensors (e.g. thermal, time-of-flight) within the environment of a hospital, to span a so called (virtual) \textit{patient zone} around the patients bed. These sensors have a low resolution, because this makes them a) cheap and thus more affordable for the hospitals and b) more energy-efficient and easy deployable, because they need no extra energy-supply.

Of course, this specific setup comes at a cost. Because of the low resolution, the recognition task itself becomes more challenging due to low-quality data. Also the spatial environment might vary for different locations. A patient room can be arranged in various ways, so that sensor data for the same scene can look differently. Therefore, one-fits-all solution for a recognition system is infeasible and a rather generalised approach is required to accommodate multiple environmental settings.
Also the fact that we are in a clinical environment plays a crucial role, when it comes to recording training data for the system. In such an environment we cannot expect to have much training data, what makes the recognition task even harder. 
 
Besides the information extracted from the sensor data, domain knowledge is also available. 
This can be provided by experts in terms of best practises for states or by semi-formal specifications for high-level activities. 
The challenging aspect here is, that the best practices for states differ between hospitals due to specific setups, personnel situation, etc., whereas the semi-formal specifications of activities stay the same, as they are specific in laws and global rules. 
Also the domain knowledge is not readily available as well-structured and annotated data, rather, such information is often possessed by respective people. 
Therefore, a formalised version of such knowledge is required to be operable by computers.

Another aspect is, that the overall system must be easily usable and intelligible for users with no technical background such as hospital staff. 
Moreover, the before mentioned heterogeneity needs to be addressed as well. 
For example, the hospital staff must be enabled to customise the system by redefining processes or incorporating additional architectural information of the hospital, such that the system is easily modified to fit the local requirements.

To summarise, our goal is to provide a low-training system, which 
a)~is able to make use of different sensor modalities with low resolution, 
b)~can make use of additional domain knowledge,
and c)~is easily accessible and customizable for non-technical users,   
This system can then serve as a baseline for further clinical use cases, such as, the entry and exit of persons into the patient zone (a virtual monitoring  zone), fall detection, bed preparation, pressure sore risks or hands-disinfection recognition for hygiene management.

\section{Patient monitoring as a use case for NeSy-AI}
\label{sec:nesy_pm}
As explained in Section~\ref{sec:patient_monitoring}, we want to integrate domain background knowledge with sensor data to infer corresponding states and activities. 

\begin{figure}
	\centering
	\includegraphics[width=1\linewidth]{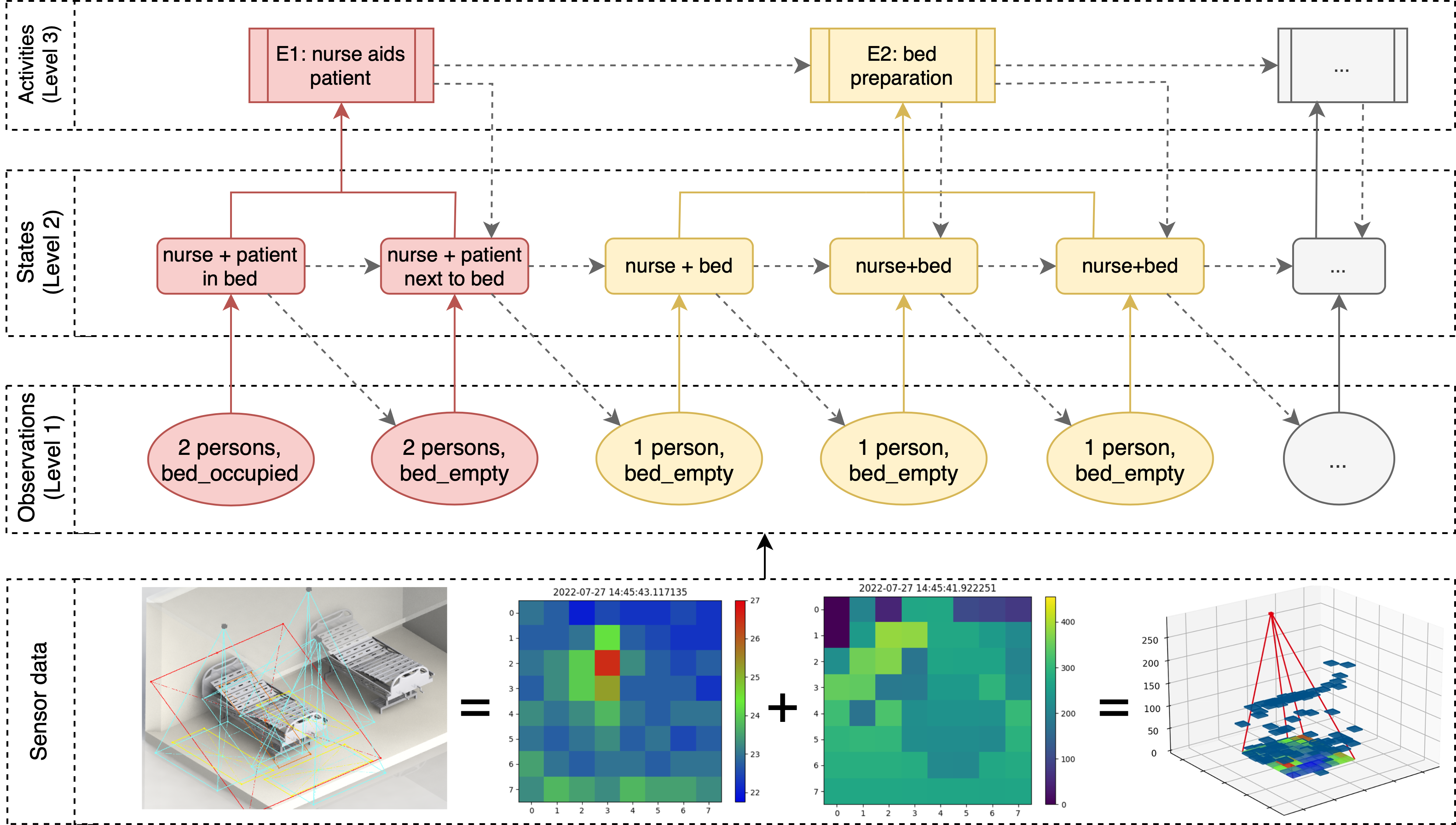}
	\caption{The different levels of the patient monitoring process.}
	\label{fig:arch}
\end{figure}

To account for the different types of domain knowledge (as outlined in Section~\ref{sec:patient_monitoring}) we divide the overall monitoring process into multiple levels, as shown in the Figure \ref{fig:arch}. 

In the first level, we extract abstract information from the observations available through the sensor data.
Here, an observation corresponds to the information that can be drawn directly from the respective sensor data, for instance, “There are two persons and an occupied bed”.

At the second level, we infer semantic states derived from the observations. 
They correspond to fine-grained semantic states, which can be inferred from observations by utilising the additional domain knowledge, specifying the concrete setup within the hospital. 
Referring back to example observation “There are two persons and an occupied bed”, we can derive the state as “Nurse + Patient” using the (domain) knowledge of being in a hospital environment.

In the third level, high-level activities are detected, which are defined as a sequence of states. 
Again, background knowledge is used at this level to infer activities. 
Lets continue our example and have a look at the next state in Figure \ref{fig:arch}. 
From the given state sequence we can conclude the activity “nurse aids patient” and that the next activity could be “bed preparation”. 
Here, domain knowledge from nursing science is utilised to map sequential states into more complex high-level activities and to predict future activities as well.

As the domain knowledge is either informal or semi-formal, a knowledge model is required to formalise it such that it can be used in AI methods. 
As described in Section~\ref{sec:NeSy}, symbolic AI is the appropriate methodology for knowledge-modelling in an interpretable and formalized manner. 
Moreover, as mentioned already, to analyse sensor data, we will use neural networks as they are proven to be robust against noise and inconsistencies in raw data. 

As this is in line with the overall goal of NeSy-AI, we argue that it should be applied to patient monitoring. 
To further underpin our argumentation, we describe some specific challenges which arise from the requirements we identified in Section~\ref{sec:patient_monitoring} and we will also explain how NeSy-AI can be used to address them.

\paragraph{Sample efficiency:}
As outlined before, neural networks are well established for low-level perception. 
However, they rely on high quantity of underlying training data. 
Therefore, in order for such networks to be applicable, samples for each possible activity with its corresponding states and observations need to be recorded as well. 
This conflicts our pivotal low-training requirement (Section~\ref{sec:patient_monitoring}).
Since additional background knowledge is available, we should utilise it to address the problem. 
In this regard, several NeSy-AI approaches are available: Logic Tensor Networks (LTNs) \cite{donadello_logic_2017} for example, integrate neural networks with first-order fuzzy logic to learn efficiently while requiring fewer data samples. 
In our context, we could potentially use fuzzy logic to model domain knowledge to achieve similar results. 
Similarly DeepProbLog \cite{manhaeve_deepproblog_2018} take one step further. 
Preliminary experiments\footnote{\url{https://github.com/vac-mmis/Neuro-SymbolicArtificialIntelligenceforPatientMonitoring}} have shown that DeepProbLog can even be used to reduce the complexity of the learning task itself, and thus the amount of training data required. 
Here, the domain knowledge can be used to model interdependence among observations (e.g., number of persons or bed), states (e.g., nurse + bed or nurse + bed + patient) and the activities (e.g., changing bed sheets or patient being asleep) associated. 
Therefore, by logically modelling the domain knowledge and integrating it with the neural networks, we can restrict the learning task to observations only. 
Thus, the respective states and activities could be inferred from the logical model while learning the low-level observations from the raw sensor data by neural networks.

\paragraph{Uncertainty:}
One of the most prominent frameworks for high-level reasoning is logic, however, it is restricted by crisp (e.g. Boolean) decisions. 
This becomes problematic in the presence of uncertainty as in noisy sensor data and, therefore, needs to be accounted for in the overall monitoring process. 
The corresponding research in the NeSy-AI field \cite{besold_neural-symbolic_2017, manhaeve_deepproblog_2018, de_raedt_neuro-symbolic_2019} argue in favour of integrating probability as its third component. 
With ProbLog \cite{de2007problog,fierens_inference_2015} for example, a dedicated language was designed to assign Prolog clauses with their probability of being true. 
In similar context, Computational State Space Models (CSSMs) use probabilistic filters for doing the inference and reasoning, and they are proven to handle problems with big state spaces well \cite{kruger_computational_2014}. 
The main problem with both approaches is, that they don't integrate with neural networks to generate observations. 
To address this, DeepProbLog \cite{manhaeve_deepproblog_2018} uses so-called neural predicates to model the probability of a clause being true or taking a particular value by a neural network (with a Softmax output layer). 
Thus, we are able to account for uncertainty resulting from the sensor data in the further monitoring process. 
Also the system is end-to-end full differentiable, hence, the overall system can be trained on annotated data to learn the probabilities of the respective (neural and probabilistic) clauses.

\paragraph{Explainability:}
Resulting from the previously outlined requirements, i.e., enabling the non-technical users for the proposed system, explainability becomes pivotal. 
As described before, neural networks’ decision making framework acts as a blackbox and, therefore, can not be explained. 
Hence, we require a symbolic part of the system.
Several NeSy approaches (e.g., \cite{bennetot_towards_2019,bollacker_extending_2019,confalonieri_using_2021,daniels_framework_2020,diaz-rodriguez_explainable_2022}) use ontologies or knowledge graphs as their symbolic component to support the intelligibility of such systems. 
One approach \cite{daniels_framework_2020} for example, generates the knowledge graph from labeled training samples and use it in combination with WordNet to (a) enrich the labels of the (training) dataset itself and (b) refine the predictions of the neural component, such that the knowledge graph constraints are adhered. 
The \textit{X-NeSyL} \cite{diaz-rodriguez_explainable_2022} approach aims to increase the overall system performance, in addition to its explainability. 
It uses a knowledge graph (which reflects expert information and serves as a gold standard) where individual nodes are associated with a neural network to guide its learning process in such a way that the expert knowledge is not violated. 
In other words, the knowledge graph serves as a so-called feature attribution graph and reflects the information if a certain neural network feature contributes negatively or positively towards a prediction during training and reasoning. 
Another related work which can be stated here is again DeepProbLog \cite{manhaeve_deepproblog_2018}. 
By using a logic program to model the domain knowledge, the knowledge base as well as the inference process are interpretable by design.

\section{Conclusion}
With this paper, we aim at initiating further research and development in patient monitoring using NeSy-AI. 
For this, we first provided a general introduction to NeSy-AI and then introduced a process architecture for patient monitoring to account for the different levels of domain knowledge used. 
Additionally, a number of important requirements for patient monitoring were presented and based on those requirements, we outlined the potential challenges and their possible solution with NeSy-AI for patient monitoring. 
The presented work, therefore, should enable NeSy techniques to be applicable in the patient monitoring domain. 
We believe that our work will not only benefit the application domain, but also the research within the NeSy community as well. 
Moreover, the presented techniques were mostly applied to toy-examples, hence, an application to real-world problems still remains an open question in the NeSy community and thus might reveal new interesting insights in further developments.
Moreover, the presented work could be used for other assistance scenarios such as elderly care or for patients with mobility issues.

\section*{Acknowledgement}
We want to thank GWA Hygiene for providing us with the sensors as well as the additional infrastructure which is necessary for a project like this.

%
%
%
\bibliographystyle{splncs04}
\bibliography{PharML23}
\end{document}